# Recent advances in La$_2$NiMnO$_6$ Double Perovskites for various applications; Challenges and opportunities

Suresh Chandra Baral, P. Maneesha, E.G. Rini, and Somaditya Sen*
Department of Physics, Indian Institute of Technology Indore, Indore, 453552, India
* Corresponding author: sens@iiti.ac.in

**Abstract**

Double perovskites R$_2$NiMnO$_6$ (R= Rare earth element) (RNMO) are a significant class of materials owing to their Multifunctional properties with the structural modifications. In particular, multifunctional double perovskite oxides La$_2$NiMnO$_6$ (LNMO) which possess both electric and magnetic orderings, chemical flexibility, versatility, and indispensable properties like high ferromagnetic curie temperature, high absorption rates, dielectrics, etc. have drawn a lot of attention due their rich physics and diverse applications in various technology. This justifies the intense research going on in this class of materials, and the keen interest they are subject to both the fundamental and practical side. In the view of the demands of this material in lead free perovskite solar cells, photocatalytic degradation of organic dyes, clean hydrogen production, electric tuneable devices, Fuel cells, gas sensing and in Biomedical applications, there is a need for an overview of all the literature so far, the ongoing research and the future prospective. This review summarised all the Physical and Structural properties of LNMO such as electric, magnetic, catalytic, and dielectrics properties with their underlying mechanisms. The mechanism of ordering at the B-site has always been a topic of debate for most of the double perovskites. This review also tries to give an insight to detailed background of the ordered and disordered phase formation at the B-site. The effect of Magnetic field as well as temperature on the electronic structure of this material are also described. The discussion also includes how different synthesis process and annealing condition can result different phases and properties of these materials. Various technique such as XRD, neutron diffraction method, Raman studies are included to give a better view towards the structure corelated functionalities. The modification of either A-site/B-site or both A-site and B-site can tune the properties of the material for desired functionalities are also described. In the end, a comparative analysis of LNMO has also included to give better overview among other double perovskite oxides for application point of view. This review article provides an insight into the scope of studies in LNMO material for exploring unexposed properties in new material research and to identify areas of future investigation of the materials in the double perovskite family.

## 1. INTRODUCTION

The growing energy demand and rapid climate changes in recent times have forced scientists to come up with eco-friendly solutions for clean energy applications. Several approaches have been proposed to produce multifunctional materials which possess both electric and magnetic orderings, has largely been driven by the growing need to lower the consumption of power and provide additional functionality for next-generation electronic devices.[1].The best option for these types of properties can be perovskite materials. Due to their physical properties, Perovskite materials have been extensively investigated for both theoretical modelling and practical applications. Potential applications include sensors, catalyst electrodes, certain types of fuel cells, solar cells, memory devices, spintronics, and lasers [2]–[11].

This unusual range of interesting properties arises due to its exceptional compositional and structural flexibility. cubic perovskite material has a chemical formula $ABX_3$. A site generally occupied by a twelve-coordinated large cation, B site with an octahedral coordinated smaller cation, and X is an anion, most commonly oxygen. Essentially, corner-shared network is formed with the octahedral B cation and the voids are filled by A cation. Furthermore, the $BX_6$ octahedra allow expansion/contraction and tilting to compensate for smaller ionic sizes. Aside from that, the octahedra may change shape. the cations may move out of their mean positions when there is an electronic instability. On any of the three sites of the structure, partially substituting or creating vacancies is also possible. Unlike traditional semiconductors, perovskites are also flexible material. Thus almost every periodic table elements can be accommodated in the structure one way or another[12], [13]. Precisely tailoring the properties of perovskites involves partial cation substitutions. Substitutions can occur at either site A or B to different extent. However, replacing half of the B-site cations by another cation has attracted many attentions. The structure of the two cations at the B site can ordered or disordered. But the B site ordered structure results in a double perovskite material with a chemical formula $A_2B'B''X_6$ [14]–[16]. More details are already given in our previous literature [17].

In more detail, double perovskites are formed by replacing exactly half of the B-site cations with another B' cations and achieving rock salt ordering between them. Or we can say that double perovskites have the structure of two single perovskites coupled together by two different transition metals (TM) at the conventional B-sites [18].

Many research groups around the world have studied the properties such as ferroelectricity, multiferroicity, and dielectric, magnetic as well as luminescent properties of double perovskites [19]–

[25]. Figure 1 shows a graphical overview of various properties associated with Double perovskite oxides.

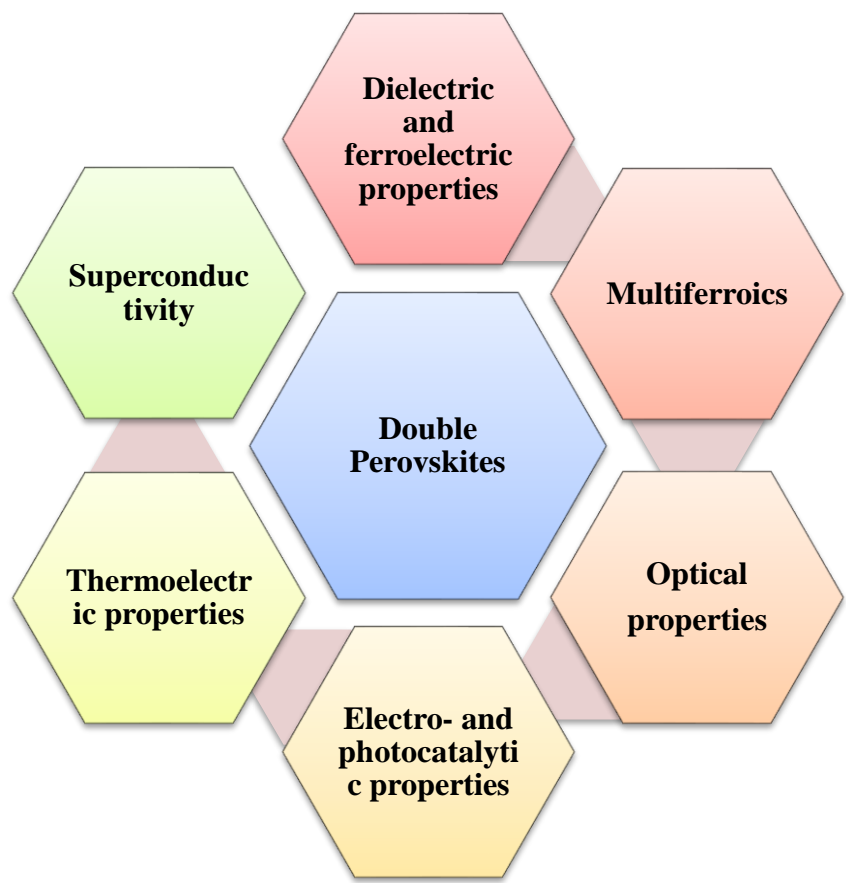

Figure 1: Various properties of Double perovskite oxides.

Among all the Double perovskite oxides, Double perovskites $R_2NiMnO_6$ (R= Rare earth element) (RNMO) are of materials with prime importance due to their multifunctional properties. In particular, the double perovskite $La_2NiMnO_6$ (LNMO) is among the few low-cost, single-material platforms that can be manipulated by magnetic or electric fields to tune various properties [26]–[29]. It has many practical applications like in lead free perovskite solar cells, photocatalytic degradation of organic dyes, clean hydrogen production, electric tuneable devices, Fuel cells, gas sensing and in Biomedical applications.

2. **Crystal structure**

For many generations of students, it is believed that in order to understand "materials science" the "structure-property" relation is the main key. However, the former is not always true and the perovskite $La_2NiMnO_6$ is an outstanding example of that [8].

The general chemical formula for perovskite oxides is $ABO_3$, where A site generally occupied by a twelve-coordinated large cation, B site with an octahedral coordinated smaller cation, and X is an anion, most commonly oxygen [30]–[35]. Different perovskite analogues can form by ordering of two different cations at the A site or the B site. The double perovskite structure is one where two kinds of cations (B and B′) occupy the B site in an ordered fashion such that the structure consists of alternating corner-connected $BO_6$ and $B'O_6$ octahedra. An alternative way of visualizing the structure is to think of it as two interpenetrating face-centred cubic lattice of B and B′ cations.

### 2.1. B-site ordering

Now one can ask the question how this ordering takes place in the double perovskites?

In order to Visualise this Kleibeuker et al.[36] have demonstrated different ways to get perfect *B*-site ordering in double perovskite thin films. To visualize the formation of ordered B-sites, they first discussed the effect of 'in-plane' strain on non-distorted $BO_6$ octahedra for different substrate orientation. There are four reduced *in-plane* B-O bonds ($d_{B\text{-}O}$) under compressive *in-plane* strain and two elongated *out-of-plane* B-O bonds ($d_{B\text{-}O}$) along (001) orientation. But along (111)-orientations, all six $d_{B\text{-}O}$ stretch to maintain the unit cell volume. then octahedral rotations takes place [31]. a zig-zag random pattern of the -O-*B*-O-*B*- chain is formed due to the rotations of the octahedra (Figures 2c and d). the rotation axes are either parallel or perpendicular to the substrate surface along (001)-oriented films (Figure 1c). the effect of strain is independent of the type of rotation (in-phase or anti-phase) of $BO_6$ octahedra. one set of $BO_6$ octahedra tilts towards the substrate surface plane (the *in-plane* direction), while the nearest neighbouring set of $BO_6$ octahedra tilts toward the substrate surface normal (the *out-of-plane* direction) For (111)-oriented films. Due to the effect of *in-plane* strain on the two sets of octahedra different *B*-site cages are formed orientating in different direction and volume. the magnitude of the strain and octahedral tilting determine the octahedral differences. The volume difference between the two octahedra is about 0.3–3%. As Both the octahedra tilt in same angle with respect to the *out-of-plane* direction thus applying *in-plane* strain on these octahedra results in a change of shape and volume, not taking cation displacements into account. The two *B*-site cages arranged in a rock salt fashion results in ordered $A_2BB'O_6$ double perovskites [37].

#### 2.1.1. Degree of order and disorder

The degree of *B*-site cation order can be quantified as a first approximation by a simple long range order parameter, defined as

$$S = 2g_B - 1$$

where $g_B$ is the occupancy of a *B*-site cation at its correct site. The completely ordered double perovskite corresponds to $S = 1$ and the completely disordered one to $S = 0$. However, partial cation ordering is also possible. What is more, the range of cation order in a single compound can vary widely due to different synthesis conditions, which allows for the control of cation order in some of the $A_2BB'O_6$ compounds. For LNMO the value of S lies between 0.8 to 1 confirming that the material can have some finite degree of disorderness in it depending upon different synthesis conditions.

### 2.1.2. Charge and size differences

The magnitude of the charge and ionic size difference between the B and B′ cations play a significant role in the ordering of these cations. A larger difference in either charge or size results in high degree of ordering of the B and B′ sites which can be observed by the appearance of the odd hkl reflections in the X-ray diffraction pattern such as the (111) peak.

#### 2.1.2.1. Effect of Cation size

As mentioned above, the ionic size difference between the B and B′ cations play a significant role in the ordering of these cations. So, it is necessary to keep in mind that the stability of the compound Is the most desirable things of the perovskite structure. Speaking about stability, the Goldschmidt tolerance factor (t.) is an important parameter to describe the stable structure of Perovskites. It is ratio of ionic radii of cations and anions that describes how well the ions that make up the perovskite structure fit together [38].

$$t = \frac{R_A + R_O}{\sqrt{2}(R_B + R_O)} \quad (1)$$

$$\mu = \frac{R_B}{R_O} \quad (2)$$

where $R_A$, $R_B$, and $R_O$ represent ionic radii of A, B cation and Oxygen respectively[39].

The tolerance factor (t) for double perovskites materials can be calculated using an average of the B and B′ ionic radii. the A site is a diamagnetic alkali earth metal; the B site is a, 3d transition metal; the B′ cation is also a 3d transition metal. Double perovskites are a great playground for studies of magnetism due to the fact that one can place a wide variety of different cation combinations at the B and B′ sites and investigate different magnetism as a function of different combination of d electron count.

$$t = \frac{R_A + R_O}{\sqrt{2}\,[(R_B + R_{B'})/2 + R_O]} \quad (3)$$

The value of the tolerance factor lies between 0.88 and 1.1 for a stable crystal structure of perovskite. It is usually expected that perovskite is stable if 't' lies within the specified range, but it is also seen that perovskite is not stable even if 't' is in the range of 0.8–0.9 [40]. An additional consideration for perovskite formation is taken into account, with the octahedral factor (µ). It is used to determine the distortion and stability of the perovskite structure. The perovskite is stabilized for an octahedral factor ranging from 0.45 to 0.89 [41], [42]. For LNMO the tolerance factor lies between 0.84-1. the tolerance factor for rhombohedral phase was calculated to be 0.89 and monoclinic phase it is 0.84 [6].

A tolerance factor equal to one is ideal to stabilize the ideal cubic perovskite with the space group $Fm\bar{3}m$. When the tolerance factor is less than 1, the perovskite framework compensates for the size mismatch via tilting of the BO6 octahedra. As the tolerance factor decreases, the amount of octahedral tilting increases, results in a lowering of symmetry from $Fm\bar{3}m$.

Octahedral tilting is usually described by the Glazer notation which is composed of three letters with by a superscript for each letter [43], [44]. The letters specify the magnitude of the tilt along the three crystallographic axes, and the superscripts designate the relative direction of tilts in successive layers along each axis. A superscript of "0" indicates no tilting, "+" indicates in-phase tilt, and "−" indicates out-of-phase tilt.

In LNMO, the glazer notations for Monoclinic $P2_1/n$ (a− a − c +) the space group have octahedral tilts that are out-of-phase about the a and b axes and in-phase about the c axis [65,114,179,180]. Lattice parameters of the monoclinic double perovskite unit cell are $a_m \approx a_c/\sqrt{2}$, $b_m \approx a_c/\sqrt{2}$, and $c_m \approx a_c$ where $a_m$, $b_m$, $c_m$ are monoclinic lattice parameters and $a_c$ is the cubic lattice parameter. And for Rhombohedral $R\bar{3}c$ the glazer notations are given by (a− a− a−), this means the space group have octahedral tilts that are out-of-phase about all the three co-ordinate axes.

### 2.1.2.2. Effect of Cation charge (Oxidation states)

As entropy always favours some degree of disorderness inside a material, but many of the Double perovskites were found to be very highly ordered. So, there must be other energetic terms favouring the order. It is well accepted, that the main factor causing *B*-site cation order in these compounds is the difference in the *B* and *B'* cation oxidation states, $\Delta Z_B = |Z_{B'} - Z_B|$. When $\Delta Z_B < 2$ the compounds are disordered and when $\Delta Z_B > 2$ the compounds are ordered [15], [37], [45]–[47]. this can be explained simply by considering the electrostatic repulsion between the *B*-site cations: when $\Delta Z_B$ is large, placing two highly charged *B"* cations close to each other will be energetically unfavourable compared to the ordered arrangement where the *B"* cation is surrounded only by the less charged *B'* cations. The

Madelung energy of ordering in the $A_2BB'O_6$ perovskites increases with increasing charge difference [48], [49]. When the difference in the *B*-site cation oxidation states is large enough, increase in the electrostatic repulsion overcomes the entropic term and the cations are ordered [50]. The Madelung energy of ordering has been calculated for cubic and tetragonal $A_2BB'O_6$ perovskites [51], [52]: if $\Delta Z_B = 0$, there is no change in Madelung energy with ordering and consequently there is no energy gain in cation ordering compared to a disordered case. However, the Madelung energy of ordering is proportional to $\Delta Z_B^2$, thus the energy increases quickly with $\Delta Z_B$, and the ordering energy can become a significant fraction of the total energy.

The well-ordered LNMO exhibits monoclinic ($P2_1/n$) and rhombohedral ($R\bar{3}$) symmetry with $Ni^{2+}$ and $Mn^{4+}$ cations alternatively arranged at B-sites while a disordered phase exhibits an orthorhombic (*Pbnm*) and rhombohedral ($R\bar{3}c$) symmetry with random distribution of $Ni^{3+}$ and $Mn^{3+}$ at the B-sites in the ideal $ABO_3$ perovskite structure [53]. While Several other reports indicate absence of any cation ordering in the LNMO lattice while others indicate fully or partially cation ordered lattices [26], [54]–[57].

With this understanding, now we can move forward how to determine the structure through experimental method.

### 2.2. Structural determination

The structure of LNMO double perovskite can be determined by different experimental method. Few are mentioned below.

#### 2.2.1. XRD Analysis

X-ray diffraction is a powerful non-destructive technique for characterizing crystalline materials. It provides information on structures, phases, preferred crystal orientations (texture), and other structural parameters, such as average grain size, crystallinity, strain, and crystal defects.

The XRD analysis shows at low temperature, LNMO double perovskite exhibit Monoclinic ($P2_1/n$) structure which converted to Rhombohedral ($R\bar{3}c$) at high temperature and the structure co exists in biphasic i.e Monoclinic ($P2_1/n$) as well as Rhombohedral($R\bar{3}c$) state in a wide range of temperature including room temperature [58]–[66]. Nasir et al.[58] have prepared Polycrystalline $La_2Ni_{1.5}Mn_{0.5}O_6$ samples by sol-gel assisted combustion method (Annealed at 1150°C air) and demonstrated X-ray diffraction that belong to a major monoclinic structural phase ($P2_1/n$), with partially disordered arrangement of Ni and Mn ions. Dass et al.[59] have demonstrated ordering of $Ni^{2+}$ and $Mn^{4+}$ ions gives ferromagnetic $Ni^{2+}$-O-$Mn^{4+}$ interactions and transforms the orthorhombic *Pbnm* space group to monoclinic $P2_1/n$ with β = 90° and the rhombohedral $R\bar{3}c$ space group to $R\bar{3}m$ or $R\bar{3}$ Synthesised by the Pechini method in Argon, air, and $O_2$ atmospheres. they have also obtained a single *P2₁/n* phase reacted under Argon atmosphere at 1350 °C and Biphasic Rhombohedral *R-3(R-3m)*, Monoclinic *P2₁/n* structure at room temperature. Sayed et al.[63] have demonstrated the Role of annealing conditions on the ferromagnetic and dielectric properties of $La_2NiMnO_6$. XRD patterns of the LNMO samples annealed in different atmospheres are shown in Fig.3. The XRD pattern clearly demonstrates the coexistence of rhombohedral and orthorhombic phases, however, no significant differences are observed for the trials with cation ordered or disordered models. that can be attributed to the smaller difference in the x-ray scattering factor of Ni and Mn. However, Kumar et al.[67] confirmed the pure-phase rhombohedral structure of LNMO with Y-doped LNMO nanoparticles corresponding to the space group $R\bar{3}c$ (SG No.: 167) with the crystallite size lying in the range of 25–16 nm. the coexistence of two phases [62.64% ($P2_1/n$) and 35.69% ($R\bar{3}$)] and [58% ($P2_1/n$) and 42% ($R\bar{3}$)] at room temperature were also demonstrated by Yang et al.[6], Kuamr et al.[68] respectively.

But Crystal structures are influenced by various properties, including ionic radius, B-site cation ordering, electronic instabilities, bond covalency, and oxygen concentration. As a result of a structural transition, it has also been detected that two phase with slightly different structures coexist. With laboratory XRD, it can be difficult to detect such phases[69]. Perovskites with $A_2BB'O_6$ crystals can also exhibit twinning, especially when their symmetry structure changes as the temperature decreases [70]. The resulting structural analysis can be affected by poor sample quality[71]. Due to the nature of these compounds, even with a high-quality sample, structural determinations can be challenging.

Unit-cell parameters can be used to determine the correct crystal system. when the structural distortions are small, double perovskites can exhibit noteworthy pseudosymmetry. The peak splitting parameter (i.e. unit-cell parameters) may not accurately disclose true structure, as the peak intensities may be affected [70], [72]–[75]. An ordinary XRD may not be able to detect such pseudosymmetry and synchrotron diffraction may be required [76].

### 2.2.2. Neutron Diffraction Analysis

Neutron diffraction experiments is another experimental technique which determine the atomic and/or magnetic structure of a material of crystalline solids, gasses, liquids or amorphous materials. The technique is similar to X-ray diffraction but X-rays interact primarily with the electron cloud surrounding each atom and diffracted x-ray intensity is therefore larger for atoms with a large atomic number (Z) than it is for atoms with a small Z. On the other hand, neutrons interact directly with the nucleus of the atom, and the contribution to the diffracted intensity is different for each isotope. As mentioned above in LNMO, the x-ray scattering factor of Ni and Mn nearly same so Neutron diffraction analysis may be used for better results.

The neutron powder diffraction data shows a mixture of the *P2₁/n* and $R\bar{3}$ structure at room temperature but completely converted to monoclinic (*P2₁/n*) at 3.5K [26]. The neutron diffraction of LNMO also confirms the coexistence of biphasic rhombohedral $R\bar{3}$ and monoclinic (*P2₁/n*) at intermediate temperature. the monoclinic phase fraction increases as the temperature is lowered [63]. The variation of unit cell parameters and phase fraction with temperature of the sample are given in table-1. Bull et al.[56] have also refined the structures LNMO using powder neutron diffraction at low- and high-temperature. At low temperature (25°C), they have showed ordered monoclinic (*P2₁/n*) and at high temperature (350-400° C) ordered rhombohedral ($R\bar{3}$) structure gave the best fittings.

TABLE 1. Unit cell parameter of LNMO-A sample at different temperatures (from powder neutron diffraction data), with permission from Ref.[63]

| Temp. (K) | Rhombohedral | | | | Monoclinic | | | | | |
|---|---|---|---|---|---|---|---|---|---|---|
| | a (Å) | c (Å) | V (Å)³ | Wt fraction (%) | a (Å) | b (Å) | c (Å) | β (°) | V (Å)³ | Wt fraction (%) |
| 350 | 5.500(1) | 13.221(2) | 346.4(1) | 68.2(2.6) | 5.430(2) | 5.491(2) | 7.738(3) | 89.80(4) | 230.7(1) | 31.8(2.2) |
| 300 | 5.497(1) | 13.215(4) | 345.9(1) | 69.0(5.0) | 5.432(3) | 5.485(3) | 7.739(4) | 89.72(6) | 230.6(2) | 31.0(4.0) |
| 230 | 5.494(1) | 13.208(3) | 345.2(1) | 65.3(2.8) | 5.430(2) | 5.488(2) | 7.734(3) | 89.85(5) | 230.4(1) | 34.5(2.4) |
| 140 | 5.488(1) | 13.216(3) | 344.7(1) | 61.7(2.7) | 5.432(2) | 5.482(2) | 7.727(3) | 89.90(5) | 230.1(1) | 38.3(2.6) |
| 80 | 5.486(1) | 13.227(4) | 344.8(1) | 53.2(3.0) | 5.433(2) | 5.483(2) | 7.719(2) | 89.80(3) | 230.0(1) | 46.8(3.0) |

### 2.2.3. Raman Analysis

Raman Spectroscopy is a non-destructive chemical analysis technique which gives detailed information about chemical structure, phase and polymorphism, crystallinity and molecular interactions. The basic principle is the interaction of light with the chemical bonds within a material. there are no Raman active vibrational modes for cubic perovskite structures however, Raman modes become active due to deviations from the ideal cubic perovskite structures.

Why there is mostly no Raman active vibrational modes for cubic perovskite structures?

When electromagnetic field interacts with the molecule's bonds, it may induce a dipole where the induced dipole moment is directly proportional to the external electric field. $P = \alpha E$ where $\alpha$ is a proportionality constant called polarizability of the molecule. Polarizability is a measure of the relative tendency of the electron cloud to be distorted from its normal shape. The change in the polarizability within the bond gives rise to Raman scattering. Scattering intensity is proportional to the square of the induced dipole moment. So, for a molecule to be Raman active, it must have a change in its polarizability. If a vibration does not greatly change the polarizability, then the polarizability derivative will be near zero, and the intensity of the Raman band will be low. As cubic perovskite is a highly symmetric structure, the Raman vibrational modes are mostly not found for a cubic perovskite structure.

For Monoclinic ($P2_1/n$) structure there are 24 ($12A_g + 12B_g$) phonon modes are Raman allowed, while for the rhombohedral $R\bar{3}$ structure there are 8($4A_g + 4E_g$) Raman modes are expected [77]. Distortions from cubic symmetry take place by rotation/tilting of the octahedron and due to displacements of the A or B cations, or B site ordering. These types of distortions are the cause for the increment in the unit cell and reduction in the symmetry which results the Raman modes that appear at the Brillouin zone centre[56]. Generally, high intense peaks were obtained at around 538 cm$^{-1}$ ($A_g$) and 683 cm$^{-1}$ ($B_{2g}$), which are due to the anti-symmetric and symmetric stretching vibration of the (Ni/Mn)O$_6$ octahedra [78]–[83]. The Raman mode around 680 ($B_{2g}$) cm$^{-1}$ have been assigned as B-O symmetric stretching vibration of oxygen in (Ni/Mn)O$_6$ octahedra. The peak around 520 cm$^{-1}$ may be due to several modes of $A_g$ and $B_g$ symmetries, involving anti-stretching and/or bending of the (Ni/Mn)O$_6$ octahedra. Kumar et al.[84] has also reported the same results. The analogy can be due to Jahn-Teller type asymmetric and symmetric stretching modes of MnO$_6$ octahedra respectively. thus, Raman analysis also consistent with the results that bulk samples are monoclinic $P2_1/n$ structure at low temperatures, coexistence of $P2_1/n$ and $R\bar{3}$ phases at room temperature, and gradual conversion to rhombohedral $R\bar{3}$ phase at high temperatures[83].

### 2.3. Effect of synthesis condition

Also, different synthesis condition can modify the structure and govern various properties in LNMO double perovskite. Table-2 shows how different synthesis condition and environment can results different structure.

Table-2 shows how different synthesis condition and environment can results different structure [85].

| Synthesis | Condition | Space group | a | b | C | β | Ref |
|---|---|---|---|---|---|---|---|
| Pechini method | 1350 K, 6 h, Air | $P2_1/n$ | 5.517 7 | 7.748 | 5.466 | 90.01 | [59] |
| Pechini method | 1623K, 12 h, Air | Pbnm | 5.501 | 5.470 | 7.751 | 90 | [86] |
| Modified nitrate Decomposition | 1373 K, 16 h, Air | $P2_1/n$ | 5.467 | 5.510 | 7.751 | 90.12 | [56] |
| Standard solid-state reaction | 1573 K, 4d, Air | Orthorhombic | 5.477 | 5.464 | 7.670 | 90 | [87] |
| Sol-gel method | 1173 K, 2d, Oxygen | 66% R3c | 5.504 | - | 12.326 | - | [54] |
| | | 34%Pbnm | 5.501 | 5.450 | 7.736 | | |
| Sol-gel method | 1673 K, | 42% R3c | 5.504 | - | 13.235 | - | [54] |

| | 12 h, Air | *42% Pbnm* | 5.503 | 5.458 | 7.727 | | |
| --- | --- | --- | --- | --- | --- | --- | --- |
| Sol-gel method | 1273 K, 12 h | *32% R3c* | 5.512 | - | 13.236 | - | [54] |
| | Arg | *68% Pbnm* | 5.512 | 5.458 | 7.739 | | |
| Sol-gel method | 700-1500 K, 6 h, Air | Monoclinic | 5.423 | 5.487 | 7.732 | | [88] |
| Glycine-nitrate | 473 K, 12 h, Air | *Orthorhombic* | 5.50 | 5.650 | 7.78 | 90 | [87] |
| Co-precipitation | 1023–1373 K, | *Monoclinic* | 5.467 | 5.510 | 7.751 | 91.12 | [6] |
| Gel combustion | Air | *69% R-3* | 5.496 | 5.475 | 13.214 | 89.66 | [63] |
| | | *31% P2$_1$/n* | 5.438 | | 7.747 | | |
| Gel combustion | O$_2$ | *50% R-3* | 5.508 | 5.473 | 13.220 | 89.48 | [63] |
| | | *50 % P2$_1$/n* | 5.468 | | 7.751 | | |
| Gel combustion | N$_2$ | *Monoclinic* | 5.482 | 5.501 | 7.770 | 89.45 | [65] |
| | | *Orthorhombic* | 5.468 | 5.506 | 12.675 | 90 | |
| | | Cubic (NiO) | 4.177 | 4.177 | 4.177 | 90 | |
| | 1273–1423 K | *P2$_1$/n* | 5.509 | 5.456 | 7.770 | 89.82 | |

## 2.4. Effects of temperature, pressure and cation vacancy

The Structural phase transitions from one tilt system to another can occur as a function of temperature, often to lower symmetries with decreasing temperature. the structural studies on bulk La$_2$NiMnO$_6$ (LNMO) provide evidence for long range ordering of Ni and Mn cations, consistent with monoclinic $P2_1/n$ ($\beta = 90°$) symmetry at low temperatures and rhombohedral $R\bar{3}$ symmetry at high temperatures. Yang et al.[6] have also explained the effect of pressure and A-site cation vacancy on the structure. They have calculated he static formation enthalpies of rhombohedral and monoclinic crystal structures per formula unit at different hydrostatic pressures via DFT within PBE at 0 K[89], [90]. the enthalpy difference (monoclinic one minus rhombohedral one) is small. This implies that the structure stability is similar for these two structures. Also, the static formation enthalpies of these two structures with A-site vacancies are extraordinarily close. That can be the reason for the co-existence of the two structures over a wide temperature range in La$_2$NiMnO$_6$ crystals. As the pressure increases, the then the rhombohedral $R\bar{3}$ phase transforms to the monoclinic $P2_1/n$ phase and vice versa.

At a high temperature, the structure of LNMO is cubic with space group $Fm\bar{3}m$. As the temperature decreases, octahedra NiO$_6$ and MnO$_6$ get rotated and transforms to the rhombohedral $R\bar{3}$ phase first. The above phase transition is of second order. then the rhombohedral $R\bar{3}$ phase transforms to the monoclinic $P2_1/n$ phase at about 650 K. but this transition is of first-order. At last, all rhombohedral phases of La$_2$NiMnO$_6$ will transform to the monoclinic phase at an extremely low temperature. Where the transition temperatures are based on the structure stability and cation vacancies.

Similarly, increase in cation vacancy also can change the monoclinic $P2_1/n$ phase to the rhombohedral $R\bar{3}$ phase which is consistent with the result obtained by Dass et al.[59].

### 2.5. Other factors affecting crystal structure

Excess amount of oxygen can also affect the crystal structure. Dass et al.[59] have studied Oxygen stoichiometry, ferromagnetism, and transport properties of $La_{2-x}NiMnO_{6+\delta}$. When the oxygen contents in the sample is less, the structure stays at monoclinic with space group $P2_1/n$ and as the oxygen contents increases it creates more and more cation vacancy in the sample converting to a mixed monoclinic and rhombohedral structure.

## 3. Electronic properties

In perovskite compounds, the electronic properties are largely determined by the B-site cations. Double perovskites are characterized by their high flexibility in combining various Elemental elements, transition metals with different transition states, lanthanoids, and actinides of different types with different oxidation states at the two *B* sites. This is why, the materials can be semiconducting, metallic, half-metallic, dielectric, ferroelectric, thermoelectric, and perhaps superconducting. The compounds containing transition metal elements often exhibit the most interesting electrical properties.

### 3.1. Electronic structure

The electrical properties of a solid are determined by the band structure. the electronic bandwidth *W*, and the intra-atomic electron-electron interactions, *U* together play important role in governing the electrical properties. when $W < U$, the electrons are localized, and when $W > U$, they can travel from one place to another. The electronic bandwidth of double perovskite largely depends on the spatial orbital overlap between the different elements. The spatial distribution of orbitals is also dependent on the crystal structure. *W* can be calculated with two parameters i.e bond length and bond angle. Due to the relative contracting nature of the 3*d* orbitals, the bandwidth in LNMO double perovskites is very low as both Manganese and nickel belongs to same 3d transition metals. Thus, there is weak overlapping of the Oxygen 2*p* orbitals. For the same reason the interelectronic repulsion in 3*d* metals is often notable, so $W < U$ is often found.

Again, matching of the two *B*-site cation orbital energies is a relatively rare occurrence, as several factors affect the energies. For transition metals, orbital energies decrease moving right in the periodic table due to the increase in nuclear charge. Similarly, crystal field splitting decreases moving right in the periodic table, due to decrease in ionic radius, and corresponding decrease in bond covalency. So orbital energy and crystal field splitting energy is different for Mn and Ni. Again, for high oxidation state there are fewer *d* electrons, and thus the relevant electrons are often found in $t_{2g}$ orbital, whereas in the high-spin 3d elements with the lower oxidation states the relevant electrons are often in $e_g$ orbitals. Thus, electron transfer is hinder in this case due to symmetry.

In ordered LNMO, nickel exists as $Ni^{2+}$ ($d^8$: $t_{2g}^6 e_g^2$, $S = 2/2$) and manganese as $Mn^{4+}$ ($d^3$: $e_{2g}^3 e_g^0$, $S = 3/2$). So, electron transfer in the *LNMO* perovskites becomes possible only in relatively rare cases with good orbital energy overlap and correct symmetries.

### 3.2. Electrical transport properties

Generally, LNMO is a ferromagnetic semiconductor with transition temperature $T_c = 280K$ [26], [58]. And in Non-stoichiometry condition Insulating has been reported [91], [92].

Ullah et al.[93] have reported Electronic, thermoelectric and magnetic properties of $La_2NiMnO_6$ and $La_2CoMnO_6$. They have calculated band structure along high symmetry points in reduced zone scheme using GGA+U [94]. LNMO shows indirect band nature for both spin up (left panel) and spin down (right panel) states. For spin up state, the minima of conduction band occur at the Γ symmetry point and the maxima of valence band are located near the Y symmetry point inside the Brillion zone. An indirect band gap (Γ–Y) is observed for spin up state with band gap value of 1.6 eV. Similarly, the minima of the conduction band are located at Γ symmetry point but the maxima of the valence band lie along Z symmetry point of Brillouin zone for spin down state. The formation of band gap along Γ–Z symmetry points is responsible for indirect band nature of compound for spin down state and the band gap energy was calculated to be 4.4 eV.

#### 3.2.1. Thermoelectric properties

To know the detail mechanism of electrical transport properties under thermal variations, it is necessary to Understand thermoelectric properties inside the material. Ullah et al.[93] have also explored the transport properties like electrical conductivity, Seebeck coefficient, thermal conductivity and power factor as function of temperature.

#### 3.2.2. Electrical conductivity (σ)

Electrical conductivity measures the flow of charge in the material. On the basis of passage of charge, material nature (conductor, insulator and semiconductor) can be easily understood. The electrical conductivity of materials varies with change of temperature. The calculated electrical conductivity per relaxation time as function of temperature (200–900 K) for LNMO is shown in the Fig. 4(a). For spin up configuration, the electrical conductivity increases as a function of temperature. At low temperature (at 200 K), the electrical conductivity was found to be to 0.50 (Ω m K)$^{-1}$. As the temperature increase to 800 K, the electrical conductivity rises up to 6.4 (Ω m K)$^{-1}$. Due to the sharp increasing property of Conductivity, LNMO can be used in Thermoelectric industries also. Nasir et al.[58] have demonstrated that resistivity of LNMO decreases with rise in temperature. Rogado et al.[26] also demonstrated the semiconducting behaviour of LNMO with resistivity ~ $10^2$ Ω cm.

### 3.2.3. Seebeck coefficient (S)

Generally, the motion of electron direct from the warmer region to the colder region. This electronic motion results in set up of electric field due to the electron-holes accretion on two sides. Ratio of voltage difference to temperature difference between these regions is termed as Seebeck coefficient (S). Seebeck coefficient determines the efficiency of thermocouples. Seebeck coefficient as function of temperature is shown in fig.4(b). the Seebeck coefficient for both compounds for spin down configuration significantly vary with temperature while for spin up configurations, no fascinating changes are observed. For spin down channel, the maximum value for the Seebeck coefficient is observed at low temperature (200 K) for both compounds. Increasing temperature induces decrease in S up to the highest temperature (800 K).

### 3.2.4. Thermal conductivity (k)

Thermal conductivity k is another important property of materials for characterizing the transparency or insulation for heat transportation. The high thermal conductivity materials are used significantly in heat sink while low thermal conductivity materials are used as thermal insulation. LNMO showed zero thermal conductivity in down spin channel and with increase in temperature the thermal conductivity increased to a maximum value at 800 K.

### 3.2.5. Power factor (P = $S^2$ σ)

Ullah et al.[93] have also calculated the thermoelectric power factor for LNMO double perovskite. The thermoelectric power generation for spin up configuration increases with increase in temperature. For spin down configuration, the power factor is zero. Therefore, the compounds can be utilized for thermoelectric applications in spin up state compared to spin down state.

### 3.3. Magneto-resistance Property

Magnetoresistance (MR) effect, the magnetic field dependent change in the resistance of the materials, has attracted considerable interest due to its rich physics and technological applications [95]–[101]. according to their driving mechanisms, The magnetoresistance effects in materials are classified into several categories, namely, anisotropic magnetoresistance (AMR) [102]–[106], giant magnetoresistance (GMR)[95], [107]–[110], tunnelling magnetoresistance(TMR) [111]–[117] and colossal magnetoresistance (CMR) [118]–[124]. Generally, the colossal magnetoresistance (CMR) is observed in the manganites and the tunneling magnetoresistance (TMR) is observed in the granular perovskite. the TMR is caused by the spin-dependent tunneling between the ferromagnetic (FM) grains. TMR shows better low field response Compare to CMR.

Magnetoresistance arises due to intergranular tunnelling which is extrinsic and is dependent on the number and thickness of grain boundaries in polycrystalline samples [125]. In double perovskites tunnelling mechanism could be triggered even within a grain (intragranular) due to the presence of antiphase boundaries (APB) originating from the characteristic B/B' cation disorder in double perovskites [126]–[128]. Considering the similarity magnetic structure of $Sr_2FeMoO_6$ and near room-temperature FM transition, the LNMO can be a promising candidate of the materials with a near room-temperature TMR effect. Rogado [26] have shown magnetocapacitance and magnetoresistance properties of LNMO double perovskite at room temperature. The FM property is originated from the super exchange interaction between the alternately ordered $Ni^{2+}$ and $Mn^{4+}$ However, the inevitable Ni/Mn AS disorders will lead to the AFM $Ni^{2+}$-O2-$Ni^{2+}$ and $Mn^{4+}$-$O^{2-}$-$Mn^{4+}$ configurations[129]. the AFM interactions would exist in this ferromagnetism-based material. Further studies confirm that the AS disorders can be integrated into antiphase boundaries in the bulk LNMO, which result in an AFM coupling between the neighbouring FM domains[130], [131]. Guo et al.[132] have found that polycrystalline LNMO sample, prepared by a standard solid-state reaction method shows an evident MR effect near room-temperature. Again, the magnetic field dependent MR can be divided into two parts of the low-field MR and the high-field MR. On the basis of the magnetic and electric properties, they have suggested that the low-field MR can be attributed to the tunneling effect between the neighbouring FM domains and the high-field MR is due to the suppression of the scattering from the spin defects caused by the Ni/Mn Anti-site disorders. The resistivities showed a semiconducting-like behavior in the whole measuring temperature range. The application of magnetic field can lead to a decrease in the resistivity (see an enlarged view in the inset of Fig. 5(a)), which means the presence of a negative MR effect in this material. The temperature dependent MR under zero and 200 Oe are shown in Fig. 5(b). It can be found that the MR–T curve exhibits an obvious inflection point around $T_C$. Above $T_C$, the value of MR is small (less than 1.5%) and almost unchanged with the temperature. However, below $T_C$ the value of MR exhibits a rapid increase upon cooling. This feature strongly suggests that the MR of the LNMO is spin dependent. Thus, the near room-temperature low-field MR in FM semiconductor LNMO may provide a promising way to realize spin-valve device in one pure material.

### 3.4. Dielectric Properties and Magneto-di-electricity

Multifunctional magneto-dielectric materials whose dielectric properties depend on the applied magnetic Field have seen wide applications in the recent electronic devices such as capacitors, piezoelectric and pyroelectric transducers, capacitive magnetic field sensors, and tuneable high-frequency filters [133]–[137]. Due to the coupling between electronic and magnetic orders, these materials exhibit interesting magneto-dielectric characteristics by manipulating the electric polarization on applying magnetic field, and vice versa [86], [138], [139]. LNMO has also attended significant interest due to its giant dielectric tunability with notable relaxation behaviour at low electric fields at room temperature and for a Curie temperature (Tc) of 280 K. Many literatures have reported that the Relaxor-like dielectric behaviour of LNMO with a giant dielectric constant is due to the charge ordering of $Ni^{2+}$ and $Mn^{4+}$ at B-site [64], [140]. Dielectric properties exhibit sensitively dependence on the sintering condition, synthesis method and electrode. It has been observed that the disordered $La_2NiMnO_6$ is a unique intrinsic multi-glass system with a very large magnetodielectric coupling over a wide range of temperature including room temperature MD > 16% at 300 K [61]. Also, the high dielectric constant and relaxation behaviour might be due to the oxygen vacancies present in $La_2NiMnO_6$ [63]. LNMO possess a centre of symmetry, which rules out the origin of ionic displacement of its colossal dielectric constant for the ferroelectrics [141]. The giant dielectric constant could also stem from charge density waves, hopping charge transport, metal–insulator transition and interface effects, so it is important to study the effect of A, B and A-B site substitution on dielectric behaviour of $La_2NiMnO_6$. It is expected that the substitution at A and B site cation will give anti-site or anti-site defects, ionic shift, change in tolerance factor, change in bond length and bond angle in $La_2NiMnO_6$. The introduction of anti-site defects in partially substituted $La_2NiMnO_6$ could give nano polar regions (NPR) which might be responsible for dielectric relaxation under the influence of field. Therefore, it will be interesting to determine the contribution of various dielectric relaxations and conduction mechanism of charge carriers in A-site, B-site and A-B site substituted $La_2NiMnO_6$ [142].

### 3.5. Ferroelectrics

According to the theory of hybrid improper ferroelectricity, combination of oxygen octahedral rotations in three-dimensional structure and layered ion ordering may cause opposite antipolar motions of A-site ions in perovskite material [126],[127]. Many reports have been reported LNMO showing Ferroelectric properties[86], [145], [146]. Zhao et al. [147] have reported $R_2NiMnO_6/La_2NiMnO_6$ superlattices along -b direction where R is a rare-earth ion—that exhibit an electrical polarization and strong ferromagnetic order near room temperature. Also Sun et al.[143] have also demonstrated the ferroelectric properties of $Pr_2NiMnO_6/La_2NiMnO_6$ superlattice and find that only the [111] order has spontaneous polarization of 2.91μC/cm$^2$ (along the -**b** direction).Takahashi et al.[148] have investigated the possibility of A-site driven ferroelectricity in B-site ordered ferromagnetic LNMO crystals. The density functional theory calculations showed that epitaxial strain stretches the rhombohedral $La_2NiMnO_6$ crystal lattice along the [111] cubic direction, triggering a displacement of the A-site La ions in the double perovskite lattice. The lattice distortion and the A-site displacements stabilize a ferroelectric polar state in ferromagnetic $La_2NiMnO_6$ crystals. The ferroelectric state only appeared in the rhombohedral $La_2NiMnO_6$ phase, where $MnO_6$ and $NiO_6$ octahedral tilting is inhibited by the 3-fold crystal symmetry. Kumar et al.[149] have studied the Ferroelectric properties of LNMO nanoparticles the hysteresis loop has been recorded for a fixed frequency of 50 Hz with a sweeping electric filed from -20 kV/cm to 20kV/cm. The P-E loop shows weak ferroelectricity for the LNMO nanoparticles. The observed weak ferroelectricity could be understood in terms of its strong magnetic character [150]. Also, the value of polarisation seems to decrease with increase in external magnetic field, thus explaining the magneto-electric interactions.

### 3.6. Electro- and photocatalytic properties

The sunlight-driven water photolysis by the photoelectrochemical cell (PEC) is considered to be 'The Holy grail' since it is one of the cleanest approaches to address the switch-over from fossil fuels and to prevent environmental pollution. A significant number of semi-conductor materials and structures have been investigated to find the most suitable PEC devices for an efficient and sustainable $H_2$ fuel production in fuel cells. However, the overall power conversion efficiency is limited by the finite light absorption, the energy loss due to the fast charge carrier recombination rate, and the electrode degradation.so in order to overcome these limitations a new material having the desired functionality is required. Sheikh et al. [151] have demonstrated Double perovskite for photoelectrochemical catalyst properties with $BiFeO_3/La_2NiMnO_6$ where the light-harvesting property has been evaluated with BFO/LNMO photoanodes in PEC water photolysis. Several others have also reported that LNMO is a promising material for oxygen evolution reaction (OER) in energy storage devices [152]–[155]. Methane is currently the second most important greenhouse gas emitted from human activities (carbon dioxide is first). But on a per molecule basis, methane is a much more effective greenhouse gas than $CO_2$. The potential contribution of methane to the global greenhouse effect has proven to be about 20 times stronger than carbon dioxide. For complete elimination of methane emission from natural gas engines and power plants, Ding et al. [156]have reported that LNMO–MgO composite oxide can be used for higher catalytic combustion of methane. Also, Tang et al. [157] have demonstrated Mesoporous double-perovskite LNMO for photothermal synergistic degradation of gaseous toluene and Shi et al. [158] also showed perovskite-type oxide catalysts for ethane oxychlorination. so LNMO can be a best alternative for fossil fuel production in the future.

### 3.7. Magnetic properties

LNMO can be manipulated by magnetic or electric fields to tune spin, electric charge, and dielectric properties. Near room temperature, this behaviour occurs, which indicates that LNMO is also a potential candidate for practical spintronic applications [26], [28], [159] .

### 3.7.1. Magnetism in ordered Structure.

$La_2NiMnO_6$ is a ferromagnet Semiconductor ($Tc \approx 280$ K) [63], [160] with ordered $Ni^{2+}$ ($d^8$:$t_{2g}^6 e_g^2$, $S = 2/2$) and $Mn^{4+}$ ($d^3$:$e_{2g}^3 e_g^0$, $S = 3/2$) ions occupying the metal (M) centres of corner-sharing $MO_6$ octahedra in a distorted perovskite structure, while the end components $LaMnO_3$(LMO) and $LaNiO_3$(LNO) are of antiferromagnetic and paramagnetic nature, respectively [161], [162].

Based on Kanamori-Goodenough rules [163], the ferromagnetic order derives from the presence of 180° $Ni^{2+}$–O–$Mn^{4+}$ super exchange bonding between an empty $Mn^{4+}$ $e_g$ orbital and a half-filled $d$ orbital on a neighbouring $Ni^{2+}$ site, which has been theoretically predicted and experimentally reported [6], [26], [59], [87], [148], [154], [164]–[169].

#### 3.7.1.1. Role of Super exchange interactions.

Super exchange is the magnetic interactions between two magnetic cations through virtual electron transfer mediated by a nonmagnetic cation. A set of semi-empirical rules were developed by Goodenough and Kanamori called the Goodenough-Kanamori rules that successfully rationalized the sign of the magnetic exchange between 3d transition metal cations bridged by an anion [170], [171]In general, the rules state that super exchange interactions are antiferromagnetic if the virtual electron transfer occurs between two half-filled orbitals, and ferromagnetic if the virtual electron transfer occur between a half-filled and an empty orbital.

Super exchange between $e_g$ orbitals of cations that are in a linear bond through a ligand occurs via virtual electron transfer with electrons of opposite spin to the respective d orbitals. This typically means that 180° super exchange is antiferromagnetic between cations that both have empty $e_g$ orbitals and ferromagnetic between a cation that has a half-filled $e_g$ orbital and one with an empty $e_g$ orbital. In addition, cation-anion-anion-cation long range super exchange can happen but typically the magnitude is significantly reduced as the number of orbitals involved in the super exchange pathway increases [171]. The sign of the interactions is similar to the short-range interactions. Similarly, Magnetic interaction between octahedral site transition-metal ions via an oxygen ion according to the Kanamori-Goodenough rule is through 180° cation-anion-cation interactions [172].

### 3.7.2. Magnetism in dis-ordered Structure.

It is also reported that LNMO contains two ferromagnetic phases due to the existence of inhomogeneities in the sample [59], [87], [173], [174]. The $Ni^{2+}$–$O^{2-}$–$Mn^{4+}$ super exchange leads to high temperature ferromagnetic phase, $T_C$ ~266 K while the $Ni^{3+}$–$O^{2-}$–$Mn^{3+}$ exchange results in a low temperature ferromagnetic phase at ~100 K [59], [87], [175]. Choudhry et al.[61] contradicted this result and described the low temperature magnetic transition as a glassy state due to the presence of disorders with $Ni^{2+}$–$O^{2-}$–$Ni^{2+}$ and $Mn^{4+}$–$O^{2-}$–$Mn^{4+}$ antiferromagnetic exchanges. Singh et al.[175] reported that $T_C$ is extremely sensitive to the Ni/Mn atomic ordering. Disordered LNMO films exhibit lowered $T_C$ ~138 K due to $Ni^{3+}$–$O^{2-}$–$Mn^{3+}$ but $T_c$ increases to 270 K with ordering, which was attributed to the $Ni^{2+}$–$O^{2-}$–$Mn^{4+}$ configuration. However, Pal et al.[176] contradicted this result and emphasized that $T_C$ and valency of Ni/Mn ions are insensitive to the atomic ordering. Nasir et al.[177] prepared LNMO samples with invariant valence states of $Ni^{2+}$ and $Mn^{4+}$. These samples revealed a single high temperature ferromagnetic transition ($TC$ ~280 K) due to $Ni^{2+}$–$O^{2-}$–$Mn^{4+}$ super exchange, which remains invariant with annealing temperature. Theoretical saturation magnetization of LNMO is 5 $\mu$B/f.u. Wang *et al.*[178] reported a saturation magnetization of 2.2 $\mu$B/f.u. for LNMO. Pal *et al.* [176] reported a saturation magnetization of 3.2 $\mu$B/f.u. and 1.2 $\mu$B/f.u for an ordered and disordered LNMO, respectively. Antiferromagnetic coupling caused the reduction of saturation moment due to the anti-site disorders, induced by the partial interexchange between the Ni and Mn sites. This type of anti-site disorders naturally occurs in double perovskites, which strongly affects their magnetic and electric properties. Goodenough *et al.*[60], [163] observed antiferromagnetic antiphase boundaries (APBs) due to high degree of anti-site disorder. Thus, anti-site disorder result in an antiferromagnetic coupling because of $Ni^{2+}$–$O^{2-}$–$Ni^{2+}$ and $Mn^{4+}$–$O^{2-}$–$Mn^{4+}$ interactions, thereby reducing magnetization.

In the absence of long-range ordering of the Ni and Mn, the most predominant interactions in $La_2NiMnO_6$ arising from the Ni–O–Ni and Mn–O–Mn links are antiferromagnetic. As a result, the disordered octahedral cations are expected to interact ferromagnetically via vibronic super-exchange coupling [57]. Furthermore, due to oxygen non-stoichiometry, the orthorhombic distortion of the cation disordered lattice depends on annealing conditions [27], [54]

## 4. Applications
### 4.1. Application in Solar cells

Photovoltaic applications have gained popularity due to the increase in solar power demand. New perovskites have been developed containing inorganic and organic lead halide with long Carrier diffusion lengths, high absorption coefficients, and appropriate surfaces. The appropriate band-gap, greater charge carrier mobility and solution-processable fabrication technique make these material promising candidate for Solar cell applications [179]–[181]. Perovskite solar cells (PSCs) has achieved an extraordinary high power conversion efficiency starting from 3.8% in 2009 to approximately 25.5% to date, becoming a promising alternative of traditional silicon solar cells [182], [183]. Though PSCs has gone through many significant advancements in hybrid lead halide perovskite materials, the commercialization of this technology is still limited due to some scientific challenges. The contains of lead in these materials is the major serious drawback for the environment and human health as it is very toxic in nature. the stability of these PSCs is not very good in devise platform. This is the reason for the lead halide perovskite solar cell (PSC) not overcoming silicon-governed photovoltaic market. Considering this severe issue into account, it becomes inevitable to investigate the alternative of lead-based perovskite which will offer superior Photoconversion efficiency and stability [184]–[186].

The appropriate band-gap of $La_2NiMnO_6$ (LNMO) makes these material to get attention for photovoltaic applications [67], [187]. the easiest material synthesis condition and chemical processing for the fabrication of double perovskite layer are additional advantages taking LNMO in consideration [85]. Kitamura et al. have found the optical band-gap of 1.5eV for LNMO epitaxial films observed from optical spectra, which are similar to $CH_3NH_3PbI_3$ with a band-gap of 1.5eV [188], [189]. Also, Zhang et al. demonstrated the sol-gel method for the deposition of polycrystalline LNMO [190]. at room temperature, LNMO has the disordered orthorhombic and ordered monoclinic crystal structures and for the first time in 2016, Lan et al. [191] reported their experimental and theoretical studies to explore the probable application of double perovskite LNMO in solar cells. The experimental results exhibited the monoclinic and rhombohedral structure of LNMO having a band-gap of 1.4eV and 1.2eV respectively. The obtained results demonstrated that the monoclinic structure of double perovskite LNMO is superior contender compare to the rhombohedral structure for photovoltaic applications. The double perovskite LNMO has two ferroelectric curie transition temperatures (~60 K and ~285 K) [67], [192]. But this material doesn't not show ferroelectric properties at room temperature instead, revealed high dielectric constant that will help to separate photo-generated charge carriers through dielectric screening. On attempting for the development of lead-free inorganic double perovskite materials, Sheikh et al. [193] demonstrated about $Ln_2NiMnO_6$ (where Ln = La, Eu, Dy and Lu) for photovoltaic applications. The carrier-lifetime of this developed material is very close to silicon solar cells which is better compare to halide perovskites. Kumar et al. [194] have shown Power conversion efficiency (PCE) of inorganic lead-free double perovskite $La_2NiMnO_6$ photovoltaic material to be 15.42% with

FF = 55.57%, Jsc = 40.64 mA/cm$^2$ and Voc = 0.6828 V using the SCAPS simulation and first principle density functional theory (DFT) calculations.

Pal et el. [195]have explored the Defect and interface engineering of highly efficient La$_2$NiMnO$_6$ planar perovskite solar cell theoretically. The simulated device is in A planar heterojunction device architecture ITO/TiO$_2$/LNMO/Spiro-OMeTAD. They have demonstrated that the PCE about 22% can be achieved if the shallow and mid gap trap densities are confined to $10^{15}$/cm$^3$ when the active layer thickness is in the range 400–550 nm. The highest conversion efficiency of a single junction PSC achieved till date is 25.5% [182]. So by employing LNMO as an active material having a direct band gap of 1.4eV that comes very close to the Shockley-Queisser bandgap value of 1.34eV for single junction solar cells, the current requirement can be achieved [196], [197]. They have also studied illumination intensity dependent JSC and VOC and noticed that the PSC has an ideality factor of 2.59 and confirmed that trap assisted recombination was the dictating mechanism responsible for the losses in the structure. so further experimental studies should be carried out for Environment friendly, lead - free perovskite solar cells.

### 4.2. Biological Applications

Magnetic nanoparticles (NPs) are commercially important materials as a consequence of their stability and striking magnetic property [198] and are applied widely in biological and medical areas, such as bio-separation[199], drug and gene delivery [200], quantitative immunoassay [201], and hyperthermia [202]. Recently, magnetic nanoparticles, such as CoFe$_2$O$_4$, MnFe$_2$O$_4$, Fe$_2$O$_3$, Fe$_3$O$_4$, and Fe [201], [203]–[206], have been studied mostly for biomedical applications. Double-perovskite La$_2$NiMnO$_6$ is a ferromagnetic material and In-order to be applied in biological and medical fields, La$_2$NiMnO$_6$ nanoparticles should be monodispersed to bind biomolecules. Proteins are relatively large biomolecules and usually have a tendency to accumulate at the interface between aqueous solutions and solid surfaces [207]–[211]. protein adsorption to surfaces is important in many disciplines, including biomedical engineering, biotechnology, and environmental science. Wu et al. [212] synthesized the monodispersed La$_2$NiMnO$_6$ nanoparticles by co-precipitation method, and described the magnetic properties and adsorption characteristics of bovine serum albumin (BSA). The absorbance of BSA by the NPs depended upon Temperature of annealing. The LNMO nanoparticles showed good adsorption performance in bovine serum albumin protein, and the preliminary optimized adsorption was about 219.61 mg/g that obtained for the LNMO nanoparticles annealed at 850°C.The absorbance of NPs is Summarised in Table-3. these LNMO nanoparticles are a potential carrier for large biomolecules, which will be widely used in the bio-medical field.

Table-2 Average grain size and magnetic and BSA adsorption properties of $La_2NiMnO_6$ nanoparticles [212]

| Annealing Temperature (°C) | Grain size (nm) | $M_s$ (×$10^{-3}$ emu/g) | $H_c$(Oe) | Nanoparticle mass (mg) | | BSA absorbed (mg/g) | |
|---|---|---|---|---|---|---|---|
| | | | | a | b | a | b |
| 750 | 33.9 | 1.97 | 37.5 | 5.5 | 7.8 | 51.00 | 36.84 |
| 850 | 36.5 | 3.1 | 19.9 | 6.5 | 8.2 | 189.35 | 219.61 |
| 950 | 37.9 | 1.97 | 42.3 | 5.4 | 7.2 | 51.94 | 30.24 |
| 1050 | 39.6 | 3.79 | 39.9 | 7.1 | 7.4 | 27.68 | 33.04 |

The nanoparticles were annealed at different temperatures for 2 h.

### 4.3. Application in Gas Sensing

Gas sensors are generally understood as providing a measurement of the concentration of some analyte of interest, such as CO, $CO_2$, $NO_x$, $SO_2$, without knowing the underlying approaches such as optical absorption, electrical conductivity, electrochemical (EC), and catalytic bead. However gas sensors measure a physical property of the environment around them, such as simple temperature, pressure, flow, thermal conductivity, and specific heat, or more complex properties such as heating value, super compressibility, and octane number for gaseous fuels. The latter may require capital-intensive (engines) or destructive testing, for example, via combustion, or involve the measurement of a number of parameters to serve as inputs to a correlation with the complex property of interest.

According to Biswal et al. [213], LNMO can be used as a gas sensor, which is a novel development in LNMO functionality. Gaseous environment and concentration affect the electrical conduction [214]. Hence, the space charge is formed by the molecules adhering to the electrode interface and adsorbing at grain boundaries. Through impedance spectroscopy, they have investigated the electrical conduction mechanism of the sensor with respect to its effect on space charges, Because Impedance measurement has an advantage over the D.C resistance measurement in terms of contributions from bulk, grain boundary, and space charge region. They have investigated the gas sensing property under different gaseous environments ($O_2$, $N_2$ and Argon) with varying concentration of gases at room temperature (300-350 K. The resistance values are found to increase with increase in gas concentration. For all the three gases, initially the resistance increases rapidly, and then slowly. The results show that, the position of critical pressure for Argon and $O_2$ is almost same 0.78bar. The data for $N_2$ is slightly different and its overall value is lower than either of the two gases.

## 5. Closing Remarks and Future Prospects

The subject of a review is challenging in the context of a hotly debated topic that continues to develop at a rapid pace. Therefore, rather than recapitulating current knowledge about LNMO, we thought it would be better to focus on some important key points that we believe need clarification.

1) Despite the fact that it has a band gap very close to shockley queisser limit, why there no report on experimental article related to any optical properties of LNMO?

2) So far, there are few reports on energy storage performance on LNMO particularly in supercapacitors or no reports on battery related one, although the specific capacitance is seeming to be very large. Why?

So, there is plenty of room for the further exploration of the said material in other applications such as spintronics and optoelectronics. We believe this article will give an overall picture that are already explored so far since more than many decades to the reader to create a pathway to explore it further.

**Acknowledgements**

The first author SCB acknowledges the Department of Science and Technology (DST, Govt. of India) for Inspire fellowship (IF190617). The second author MP acknowledges the Government of India for Prime Minister Fellowship (PMRF – 2101307). The corresponding authors SS acknowledges the Department of Science and Technology (DST, Govt. of India) for research grant under the AMT project (DST/TDT/AMT/2017/200), and EGR acknowledges the Department of Science and Technology (DST, Govt. of India) for financial support under the Women Scientist Scheme-A (SR/WOS-A/PM-99/2016 (G)).